\documentstyle[prb,aps,epsfig]{revtex}

         \def\Qb{{\bf Q}}

      \def\qb{{\bf q}}   
\begin{document}
\draft
\twocolumn[\hsize\textwidth\columnwidth\hsize\csname
@twocolumnfalse\endcsname

\title{Electron energy loss in carbon nanostructures}

\author{A. Rivacoba$^{1,2}$ and F. J. Garc\'{\i}a de Abajo$^2$}
\address{$^1$Materialen Fisika Saila, Kimika Fakultatea, UPV/EHU,
1072 P.K., 20080 Donostia, Spain \\
$^2$Donostia International Physics Center (DIPC) and Centro Mixto CSIC-UPV/EHU,
Aptdo. 1072, 20080 San Sebasti\'{a}n, Spain}
\date{\today}
\maketitle

\begin{abstract}
The response of fullerenes and carbon nanotubes is investigated by
representing each carbon atom by its atomic polarizability. The
polarization of each carbon atom produces an induced dipole that
is the result of the interaction with a given external field plus
the mutual interaction among carbon atoms. The polarizability is
obtained from the dielectric function of graphite after invoking
the Clausius-Mossotti relation. This formalism is applied to the
simulation of electron energy loss spectra both in fullerenes and
in carbon nanotubes. The case of broad electron beams is
considered and the loss probability is analyzed in detail as a
function of the electron deflection angle within a fully
quantum-mechanical description of the electrons. A general good
agreement with available experiments is obtained in a wide range
of probe energies between $1$ keV and $60$ keV.
\end{abstract}
] \narrowtext


\section{Introduction}
\label{Introduction}


Since the discovery of fullerenes, \cite{Kroto85} the collective
excitations of these molecules have received considerable
attention. The first electron energy loss spectra (EELS) of solid
C$_{60}$ where reported by Saito and coworkers. \cite{Saito91}
Later, EELS has been employed to study fullerenes supported on
surfaces, \cite{Gensterblum91,lucas92} single-walled fullerenes in
the gas phase, \cite{Keller92} multishell fullerenes,
\cite{Lucas94,Li94,Kociak00,Pichler01} and other carbon structures
of more exotic shapes.
\cite{Kuzuo94,Kuzuo95,Bursill94,Suenaga01,Reed01,Liu02} Along with
this experimental effort, theoretical studies of the electronic
structure and the collective excitations of these molecules have
also been carried out. \cite{Giai93,Longe93,Rubio93,Lin94} In
particular, Saito and coworkers \cite{Saito91} noted the strong
similarity between loss spectra of fullerenes and graphite, and
pointed out that this was due to the fact that the electronic
structure of both of these materials consisted of $\sigma$
orbitals along the carbon bonds and $\pi$ orbitals in the normal
direction. This idea has been exploited by Lucas {\it et al.}
\cite{Lucas94} to describe the response of fullerenes in
connection to EELS. They used a continuum model consisting of a
classical dielectric sphere that was made of a non-isotropic
dielectric function equal to that of graphite. Henrard and Lambin
\cite{Henrard96} extended that work by using the discrete-dipole
approximation \cite{Fano60,Lucas72} (DDA), where the fullerenes
were considered as a system of coupled point dipoles, and their
atomic polarizability tensor was obtained from the dielectric
function of graphite via the Clausius-Mossotti relation.

The continuum model \cite{Lucas94} works well for multishell
fullerenes, where the thickness of the target can be properly
assigned, but this is not the case of single-layer clusters, where
the thickness parameter becomes critical while it is not well
defined. However, the DDA provides a reasonable description in the
latter case.

So far, EELS electrons have been described as classical charged
particles,\cite{Henrard96} although such an approach presents some
disadvantages, connected to the fact that the energy loss spectra
depend strongly on the beam position relative to the specific
atomic positions of the carbon cluster, and in particular, the
loss probability diverges when the beam intersects any of the
atomic positions, a fact that derives from the unphysical
singularity at the origin of the dipole potential ($\sim 1/r^2$)
used in this  approach. This pathological behavior at small
distances is connected to unphysically large values of the
momentum transfer, which are included implicitly in the classical
energy loss theory. \cite{ritch81,rivacoba92} The classical theory
is then only justified for relatively distant collisions, where
the contribution of large momentum transfer is negligible. Another
drawback of a classical description of the probe derives from its
inability to account for the momentum transfer dependence of the
loss probability, a magnitude which is accessible to EELS theory
\cite{Moreau97} and  experiments \cite{Li94,Pichler01,Liu02} via
the selection of electron scattering angles: momentum-resolved
EELS has actually proved to be a useful tool for obtaining optical
and electronic properties in many cases.
\cite{Li94,Pichler01,Liu02} One more issue related to momentum
transfer (in this case with the component parallel to the electron
velocity) is the so-called recoil of the electron, which can play
a noticeable role at low electron energies like those employed in
some experiments with fullerenes in order to avoid damage of the
specimen. \cite{ritch81}

In this work, we make use of the self-energy formalism to
calculate the energy loss probability within a fully
quantum-mechanical description of the electron, while the response
of the carbon cluster is expressed in terms of the screened
interaction as calculated in the DDA via the atomic polarizability
of the carbon atoms. The energy loss probability is obtained as a
function of both the lost energy and the electron deflection angle
(i.e., the momentum transfer component normal to the electron
beam). This solves the problems that come about in a classical
description of the electron, as discussed above: divergences in
close encounters are avoided in a natural way and the contribution
of electron recoil is fully accounted for. For the electron
energies under consideration, the initial and final electron
states can be described by plane waves. This is a very realistic
approach to EELS experiments when a broad beam is employed.
Furthermore, when narrow beams of $\sim 0.5$ nm are used, one can
interpret our extended-beam results as spectra averaged over the
impact parameter, \cite{ritch81} which raises the question of the
suitability of the dipole potential $\sim 1/r^2$ near the atoms;
nevertheless close-encounter processes correspond to large
momentum transfers, and therefore, they contribute little when
calculating the spectra of electrons collected with small
scattering angle, as it is usually done in the experiments. Notice
that for classical electrons the inelastic scattering occurs up to
an impact parameter $b\le v/\omega$, where $v$ is the electron
velocity and $\omega$ is the energy loss, so that the broad beam
approach is likely to overestimate the weight of low-energy
excitations for target sizes comparable to the lateral coherence
length of the electron beam.

The screened interaction calculated in Sec.\ \ref{SecII} is
actually the Green function of Poisson's equation in the presence
of the sample, and thus, it provides the response of the fullerene
to any arbitrary external field. When we apply this theory to
simulate EELS in single layered carbon structures (Sec.\
\ref{results}), very good agreement with experiments is obtained,
indicating that the screened interaction could equally be applied
to other problems involving the response of fullerenes with
inclusion of atomic details (e.g., in STM, light absorption,
\cite{Andersen2000} van der Waals forces, \cite{GLD94} and image
states \cite{Granger2002}) . A similar approach could be also
applied to other materials such as boron-nitride structures.

Atomic units (a.u., i.e., $\hbar=m=e=1$) will be used throughout
this work, unless otherwise specified.

\section{The screened interaction of a finite atomic cluster}
\label{SecII}

We shall consider a system composed by $N$ atoms located at the positions
${\bf r}^{a}$, with $a=1, \dots N$. Each atom can be polarized so that
it behaves like an induced dipole. The polarizability tensor of atom $a$
will be denoted
${\bf\alpha}^{a}(\omega)$ in frequency space $\omega$. When an external
electric field $\bf{E}(\omega)$ is applied to the system, the induced dipole
moment ${\bf p}^{a}(\omega)$ of atom $a$ is the response to the total
field, that is, the external
field plus the field induced by the rest of the dipoles. One finds,
\begin{equation}
p^{a}_{i}=\sum_{jk} \alpha^{a}_{ij}
[E_{j}({\bf{r}}^{a},\omega)+\sum F_{jk}^{ab}p^{b}_{k}]
\label{polariza}
\end {equation}
where the indices $i$, $j$, and $k$ refer to Cartesian
coordinates.
The second term in the right hand side of Eq.\ (\ref {polariza})
involves the dipole-dipole interaction, which can be written in
terms of the relative atomic positions ${\bf r}^{ab}={\bf
r}^{a}-{\bf r}^{b}$ as
\begin{eqnarray}
    F_{jk}^{ab}&=&\frac{1}{(r^{ab})^3}[\,3\frac{r^{ab}_{j}r^{ab}_{k}}
                  {(r^{ab})^2}-\delta_{jk}], \hspace{2cm} a\neq b,
                  \nonumber \\
    &=&0, \hspace{5,15cm} a=b. \label{fff}
\end {eqnarray}
Notice that matrix $F$ is fully symmetric both in coordinate
indices $ij$ and in atomic ones $ab$. From Eq.\ (\ref{polariza}),
the dipolar moments of the system are found to be
\begin{equation}
p^{a}_{i}(\omega)=\sum_{b,jk}[M^{-1}]^{ab}_{ij}\alpha^{b}_{jk}\,E_{k}({\bf{r}}^{b},
\omega), \label{momentdipol}
\end {equation}
where $M^{-1}$ is the inverse of the $\omega$-dependent matrix
\begin{equation}
    M^{ab}_{ij}=\delta^{ab}\delta_{ij}-\alpha^{a}_{ik}F_{kj}^{ab}.
    \label{matrix}
\end{equation}

Eq.\ (\ref{momentdipol}) permits calculating the induced-dipole
moments for an arbitrary external field, from which one can in
turn obtain the induced field. In particular, this permits to
derive the so-called {\it screened interaction} of the system
$W({\bf r},{\bf r}',\omega)$, which is the solution of the Poisson
equation for a point charge at ${\bf r'}$, that is,
\begin{equation}
 \nabla^2\,W({\bf r},{\bf r'},\omega)=-4\pi\,\delta ({\bf r}-{\bf r'}).
 \label{W-0}
\label{poisson}
\end{equation}

In this case, the field ${\bf E}({\bf r})$ created by the external
probe is just the Coulomb field $1/\mid{\bf r}-{\bf r}'\mid$, and
therefore, the induced part of the screened interaction can be
straightforwardly written as
\begin{equation}
W^{\rm ind}({\bf r},{\bf
r}',\omega)=-\sum_{ab,ik}C^{ab}_{ik}(\omega)
\,\frac{(r_{i}-r_{i}^a)}{\mid{\bf r}-{\bf r}^a\mid ^3}
\,\frac{(r'_{k}-r_{k}^b)}{\mid{\bf r'}-{\bf r}^b\mid ^3}\,,
\label{screen}
\end{equation}
where
\begin{equation}
C^{ab}_{ik}(\omega)=\sum_j [M^{-1}]^{ab}_{ij}\,\alpha^{b}_{jk}.
\label{C}
\end{equation}

Like $M$ above, this matrix is symmetric in both sets of indices,
so that $W^{\rm ind}({\bf r},{\bf r}',\omega)$ is also symmetric
with respect to the spatial variables $\bf{r}$ and $\bf{r'}$.

From a practical point of view, it is important to note that the
dependence on frequency and spatial variables is separated in this
formalism: the dependence on $\omega$ is fully contained in matrix
$C$, whereas the dependence on both ${\bf r}$ and ${\bf r'}$ comes
from the last two factors in Eq. (\ref{screen}). Therefore,
although the actual dipoles have to be calculated numerically to
obtain matrix $C$, the spatial-dependent part can be dealt with
analytically in many cases. For instance, it will be useful to
rewrite (\ref{screen}) in momentum space as
\begin{eqnarray}
W^{\rm ind}({\bf r},{\bf
r}',\omega)&=&\frac{4}{(2\pi)^4}\sum_{ab,ik}C^{ab}_{ik}(\omega)
\label{screen1} \\ &\times& \int \int d{\bf p} d{\bf
q}\frac{q_{i}p_{k}}{{\bf p}^2 {\bf q}^2}e^{{\rm i}{\bf q}({\bf
r}-{\bf r}^a)}e^{{\rm i}{\bf p}({\bf r}'-{\bf r}^b)}\,. \nonumber
\end{eqnarray}

The response of the system to any external probe can be now
expressed in terms of Eq. (\ref{screen}). As an example, the
energy loss experienced by a fast classical electron is obtained
as a double integral of the screened interaction,
\cite{rivacoba92} which can be solved analytically
\cite{Ferrell87} to recover expressions (18)-(20) of Ref.\
[\onlinecite{Henrard96}].


\section{Energy loss of fast electrons in a system of coupled dipoles}

As an application of the formalism presented above, we now proceed
to compute the energy loss experienced by a broad beam of fast
electrons incoming along the $z$-axis into an atomic cluster,
described as a set of coupled induced dipoles. We study the
transition between initial and final states of the electrons,
described by plane waves $\Psi_{0}({\bf r})=V^{-1/2}e^{i{\bf
k}_{0}{\bf r}}$ and $\Psi_{f}({\bf r})=V^{-1/2}e^{i{\bf k}_{f}{\bf
r}}$, respectively, where ${\bf k}_{0}$ and ${\bf k}_{f}$ are the
initial and final momenta, and V is the normalization volume. The
transition probability can be written \cite{echeba87,jga1}
\begin{eqnarray}
P(\omega)&=&\frac{2}{v}\sum _{f}\,\int \,d{\bf r}\,d{\bf r}'\,
\Psi^\star_{f}({\bf r}')\,\Psi_{f}({\bf r})   \nonumber \\
&\times& {\rm Im}\{-W_{\rm ind}({\bf r},{\bf r}',\omega)\}
\Psi^\star_{0}({\bf r})\,\Psi_{0}({\bf r}') \,\delta (\omega -
\omega_{0f})\,, \label{selfenergy1}
\end{eqnarray}
where ${\bf v}={\bf k}_0$ is the velocity of the incoming probe,
taken to be along the positive $z$ axis,
$\omega_{0f}=\frac{1}{2}({\bf k}_{0}^2-{\bf k}_{f}^2)$ is the
transition energy, and ${\rm Im}\{W^{\rm ind}({\bf r},{\bf
r}',\omega)\}$ is the imaginary part of the screened interaction.
The sum over final states $f$ can be transformed into an integral
over the momentum transfer ${\bf Q}={\bf k}_{0}-{\bf k}_{f}$. One
finds
\begin{eqnarray}
P(\omega)&=&-\frac{2}{v}\,\frac{1}{(2 \pi )^3}\,\frac{1}{V}\,\int
d{\bf Q}\,\int \,d{\bf r}\,d{\bf r}'\, e^{-{i\bf Q}({\bf r}-{\bf
r}')} \nonumber\\ &\times&  \,{\rm Im}\{W^{\rm ind}({\bf r},{\bf
r}',\omega)\} \,\delta (\omega - {\bf Q}{\bf v}+\frac{1}{2}Q^2)\,.
\label{selfenergy2}
\end{eqnarray}
Now, inserting Eq.\ (\ref{screen1}) into this expression, one can
write
\begin{eqnarray}
P(\omega)&=&\frac{4}{\pi v}\,\sum_{ab,kj}\,{\rm
Im}\{C^{ab}_{kj}(\omega)\} \nonumber \\ &\times& \,\int \,d{\bf Q}
\,\frac{Q_{j}Q_{k}}{Q^4}\,e^{-i{\bf Q}({\bf r}^b-{\bf r}^a)}\,
\delta (\omega - {\bf Q}{\bf v}+\frac{1}{2}Q^2).
\label{selfenergy3}
\end{eqnarray}

The probability of collecting an electron after suffering an
energy loss $\omega$ and being scattered inside a circular
aperture of half-angle $\theta_{m}$ can be expressed as the sum of
the contributions of excitations characterized by $\omega$ and
$\qb$ (the energy and the component of the momentum transfer
normal to the beam direction, respectively, that is,
$\Qb=(\qb,q_z)$) as follows:
\begin{equation}
P(\omega)=-\frac{8}{v^2}\,\sum_{ab,kj}\,{\rm
Im}\{C^{ab}_{kj}(\omega)\}\,\, \Gamma^{ab}_{jk}\,, \label{pqw1}
\end{equation}
where $\Gamma^{ab}_{jk}$ is a matrix (symmetric in both $ab$ and
$ij$) that reduces to
\begin{eqnarray}
\Gamma^{aa}_{11}&=&\Gamma^{aa}_{22}=\frac{1}{2}\int_{0}^{ q_{m}}\,
dq\,\frac{q^3}{\Delta},\nonumber \\ \Gamma^{aa}_{33}&=&\int_{0}^{
q_{m}}\, dq\,\frac{qq_{z}^2}{\Delta}\,,\nonumber \\
\Gamma^{aa}_{ij}&=&0\,\,\, i\ne j\,, \label{matrix2}
\end {eqnarray}
when $a=b$. Here, the $z$ component of the momentum transfer,
$q_{z}$, and $\Delta$ are both functions of $\omega$ and $q$, that
is,
\begin{eqnarray}
q_{z}&=&v-\sqrt{v^2-2\omega-q^2}\,\label{pz} \\ \Delta &=
&(q^2+q_{z}^2)^2\sqrt{ 1-\frac{2\omega+q^2}{v^2}},
\end{eqnarray}
and $q_{m}\sim v\theta_{m}$ is the largest possible value of $q$.
The matrix elements coupling different atoms can be written as
\begin{eqnarray}
\Gamma^{ab}_{11}&=&\int_{0}^{
 q_{m}}\frac{dq}{\Delta}\,\cos[q_{z}(z^b-z^a)]\,\nonumber
\\ &\times& q^2 [q\, cos^2\theta_{ab}J_{0}(q\rho_{ab})
-\rho_{ab}^{-1}\cos(2\theta_{ab})J_{1}(q\rho_{ab})],\nonumber \\
\Gamma^{ab}_{22}&=&\int_{0}^{
 q_{m}}\frac{dq}{\Delta}\,\cos[q_{z}(z^b-z^a)]\, \nonumber \\
&\times& q^2 [q\, sin^2\theta_{ab}J_{0}(q\rho_{ab})
+\rho_{ab}^{-1}\cos(2\theta_{ab})J_{1}(q\rho_{ab})],\nonumber
\\
\Gamma^{ab}_{33}&=&\int_{0}^{
 q_{m}}\frac{dq}{\Delta}\,\cos[q_{z}(z^b-z^a)] q_{z}^2\,q\,J_
{0}(q\rho_{ab}), \nonumber \\
\Gamma^{ab}_{12}&=&-\frac{1}{2}\int_{0}^{
 q_{m}}\frac{dq}{\Delta}\,\cos[q_{z}(z^b -z^a)]\, \sin(2\theta_{
ab})\,q^3 \,J_{2}(q\rho_{ab}), \nonumber
\\
\Gamma^{ab}_{13}&=&-\int_{0}^{
 q_{m}}\frac{dq}{\Delta}\,\sin[q_{z}(z^b-z^a)]\,
\cos\theta_{ab}\,q_{z}\,q^2 \,J_{1}(q\rho_{ab}), \nonumber \\
\Gamma^{ab}_{23}&=&-\int_{0}^{
 q_{m}}\frac{dq}{\Delta}\,\sin[q_{z}(z^b-z^a)]\,
\sin\theta_{ab}\,q_{z}\,q^2 \,J_{1}(q\rho_{ab}), \nonumber
\end{eqnarray}
where $J_{n}$ is the Bessel functions of first kind and order $n$,
$\rho _{ab}=\sqrt{(x^a-x^b)^2+(y^a-y^b)^2}$, and
$\theta_{ab}=\cos^{-1}\{(x^b-x^a)/\rho _{ab}\}$.

The expression (\ref{pz}) for $q_{z}$ derives from the
$\delta-$function ensuring energy conservation in Eq.\
(\ref{selfenergy3}), where the term $\frac{1}{2}Q^2$ represents
the recoil of the electron. For large electron velocities, this
term is negligible and $q_{z}$ can be approximated by the
classical value $q_{z}=\omega/v$. Actually, Ritchie \cite{ritch81}
proved that even when all the scattered electrons were collected
the effect of this term is small for fast ($100$ keV) electrons.

The coupling between different atoms in Eq.\ (\ref{pqw1}) is
introduced via the screened interaction, where the polarization of
neighboring atoms is strongly correlated. This leads to the
existence of polarization waves that play the same role as
plasmons in extended media. The spatial extension of these waves
in the direction normal to the beam can be analyzed by studying
the dependence of the $\Gamma^{ab}_{ij}$ integrals (see above) on
the separation between dipoles in the direction normal to the
electron beam, $\rho_{ab}$, and for the sake of obtaining an
estimate one can use the classical-approach approximations
consisting in neglecting the recoil (i.e., $q_{z}=\omega/v$) and
extending the integral in $q$ from zero to infinity for $a\ne b$;
one finds
\begin{eqnarray}
\Gamma^{ab}_{33}&\rightarrow&\,\{\frac{\omega}{v}\}^2cos[\frac{\omega}{v}(z^b-z^
a)]\int_{0}^{\infty}dq\
\frac{q\,J_{0}(q\rho_{ab})}{(q^2+{\frac{\omega}{v}}^2\}^2}
\nonumber \\ &=&
\frac{\omega\rho_{ab}}{2v}\,cos[\frac{\omega}{v}(z^b-z^a)]\,K_{0}(\frac{\omega\rho_{ab}}{v}),
\label{range}
\end{eqnarray}
where $K_0$ is the modified Bessel function, and similar
expressions involving $K_m$ are found for the rest of the
$\Gamma^{ab}_{ij}$ integrals. Taking into account the asymptotic
behaviour of $K_{m}$, Eq.\ (\ref{range}) permits us to say that
the dipoles are effectively coupled up to a distance of the order
of the adiabatic length $v/\omega$, similar to the lateral
extension of the charge density associated to surface plasmons in
extended media. \cite{rivacoba00}

Equation (\ref{pqw1}) can be applied to any atomic cluster
described by the atomic polarizabilities. In the case of a system
of non-interacting dipoles [i.e, when $F_{jk}^{ab}$ is set to 0 in
Eq.\ (\ref{matrix})], assuming an isotropic polarizability,
$\alpha^{a}_{jk}=\alpha\,\delta_{jk}$, and neglecting the recoil,
the energy loss per atom probability reduces to
\begin{equation}
P(\omega)=\frac{4}{v^2}\,\sum_{a}\,{\rm
Im}\{-\alpha(\omega)\}\,\ln[1+(\frac{
 q_{m}v}{\omega})^2] \,\,. \label{limit1}
\end{equation}
This expression reproduces the energy loss probability of a system
of uncoupled dielectric spheres of radius $R$, where the
polarizability is given by
$\alpha=\frac{\epsilon-1}{\epsilon+2}R^3$, in the limit of small
momentum transfer, $qR\ll1$. \cite{Penn83,echeba87,barrera95} This
means that the suitability of this approach is restricted to
values $  q_{m}\leq 1/R$, where $R$ has to be understood as the
radius of a region containing the valence electrons. For $\sigma$
electrons one can estimate $R$ as half the distance between C
atoms in graphite (i.e., $1.42 \AA$), which limits the validity of
the present theory to $q_{m}\leq 1$ a.u. Notice that this limit is
related only to the suitability of the dipolar potential near the
origin: as one reaches the limit $qR\sim 1$, higher multipolar
excitations become more relevant than the dipolar one, and one
should rely on a multipole approach. \cite{jga2} This fact,
pointed out by Keller and Coplan, \cite{Keller92} posses another
limit to the applicability of the DDA theory.

Although it has been proved \cite{rivacoba92} that classical
energy loss expressions can be recovered from quantum-mechanical
ones by neglecting recoil and taking the $q_{m}\rightarrow\infty$
limit, such an expression is not defined in the present context,
since diagonal terms of $\Gamma$ diverge logarithmically at large
values of $q$, which is another manifestation of the unphysical
behaviour of the dipolar potential near the target-atom nuclei.


\section{The screened interaction of infinite systems: carbon nanotubes}
\label{ap-a}


The theory presented in the preceding sections can be
straightforwardly applied to finite atomic clusters, such as
fullerenes, but it is not suitable to study infinite systems. In
what follows, we will recast this theory to describe a carbon
nanotube consisting of the infinite repetition of a unit cell of
length $d$ along the direction of the $x$-axis. In order to take
advantage of the translational invariance of this system, we write
the coupling integrals $F^{ab}_{ij}$ as
\begin{equation}
F^{ab}_{ij}=\int_{-\infty}^{\infty}dq e^{iq(x^a-x^b)}
\phi^{ab}_{ij}(q)\,, \label{FT}
\end{equation}
where $\phi^{ab}_{ij}(q)$ is the Fourier transform of
$F^{ab}_{ij}$, which reduces to
\begin{eqnarray}
\phi^{ab}_{11}(q)=&&-\frac{1}{\pi} q^2K_0(\mid
q\mid\rho_{ab}),\nonumber\\ \phi^{ab}_{1i}(q)=&&-\frac{i}{\pi}q
\mid q\mid K_{1}(\mid q\mid\rho _{ab})
\frac{(r_{i}^a-r_{i}^b)}{\rho _{ab}}, \nonumber \\
\phi^{ab}_{ij}(q)=&&\frac{1}{\pi}\{q^2K_{0}(\mid q\mid\rho
_{ab})\frac{(r_{i}^a-r_{i}^b)(r_{j}^a-r_{j}^b)} {\rho _{ab}^2},
\label{fi} \\ &&-\mid q \mid K_{1}(\mid q \mid \rho
_{ab})[\frac{\delta_{ij}}{\rho _{ab}}-2
\frac{(r_{i}^a-r_{i}^b)(r_{j}^a-r_{j}^b)} {\rho _{ab}^3}]\},
\nonumber
\end{eqnarray}
with $i,j=2,3$ ($j=1$ refers to the $x$ direction) and
$\rho_{ab}=\sqrt{(y^a-y^b)^2+(z^a-z^b)^2}$. Notice that
$\phi^{ab}_{ij}(q)$ depends only on the components of the relative
position vector that are normal to the translational axis of
symmetry, and therefore, it is enough to compute it for the atoms
of a given unit cell of the tube.

For atoms located in equivalent positions of the tube (i.e., $\rho
_{ab}=0$), only the diagonal elements are involved, but obviously,
the above expressions diverge as
\begin{equation}
\phi^{ab}_{11}=-\frac{1}{2}\phi^{ab}_{22}=-\frac{1}{2}\phi^{ab}_{33}=\frac{1}{\pi}\int_{-\infty}^{\infty}dq
\, \frac{e^{-iqx}}{\mid x \mid^3}. \label{FT33}
\end{equation}
However, this divergence can be avoided on physical grounds by
excluding the self-interaction (i.e., an atom interaction with
itself) from the summation over target atoms [see Eq.\
(\ref{fff})]. In a similar way, one can write
\begin{equation}
[M^{-1}]^{ab}_{ij}=\int_{0}^{g} dq e^{iq(x^a-x^b)}
\nu^{ab}_{ij}(q) \label{FT1}
\end{equation}
where $g=2\pi/d$ is the reciprocal lattice constant and
$\nu^{ab}_{ij}(q)$ satisfies the secular equation
\begin{equation}
\frac{1}{g}\,\delta_{il}\,\delta^{ac}=\sum_{b,kj}\{\,\delta_{ij}\,\delta^{ab}-\alpha^{a}_{i
k}\,R^{ab}_{kj}(q)\}\nu^{bc}_{jl}(q)\,. \label {inverse}
\end{equation}
Here, the indices $a$, $b$, and $c$ run only over the first unit
cell and $R^{ab}_{kj}(q)$ is defined as
\begin{equation}
R^{ab}_{kj}(q)=g \sum_{n}[e^{ing(x^a-x^b)}\phi^{ab}_{kj}(q+ng)]
\label{matrix r}
\end{equation}
For $a=b$ (equivalent atoms), the $n=0$ term must be excluded from
Eq.\ (\ref{matrix r}) in order to avoid the self-interaction. This
removes the noted divergence from Eq.\ (\ref{FT33}) and allows
using finite numbers in the secular equation (\ref{inverse}).
Finally, one finds
\begin{equation}
R^{aa}_{11}(q)=-2R^{aa}_{22}(q)=-2R^{aa}_{33}(q)=\frac{4}{d^3}\,\sum_{n>0}\frac{\cos(nqd)}{n^3},
\end{equation}
which is a well defined sum.  The fact that $R^{ab}_{kj}(q)$ is a
periodic function of period $g$ has been exploited to reduce the
integration interval in Eq.\ (\ref{FT}) to the first Brillouin
zone in Eq.\ (\ref{FT1}). Once the matrix $\nu(q)$ has been
obtained, $M^{-1}$ can be computed via Eq.\ (\ref{FT1}) and the
problem is solved.


\section{Results and discussion}

\label{results}

\begin{figure}
\centerline{\scalebox{0.36}{\includegraphics{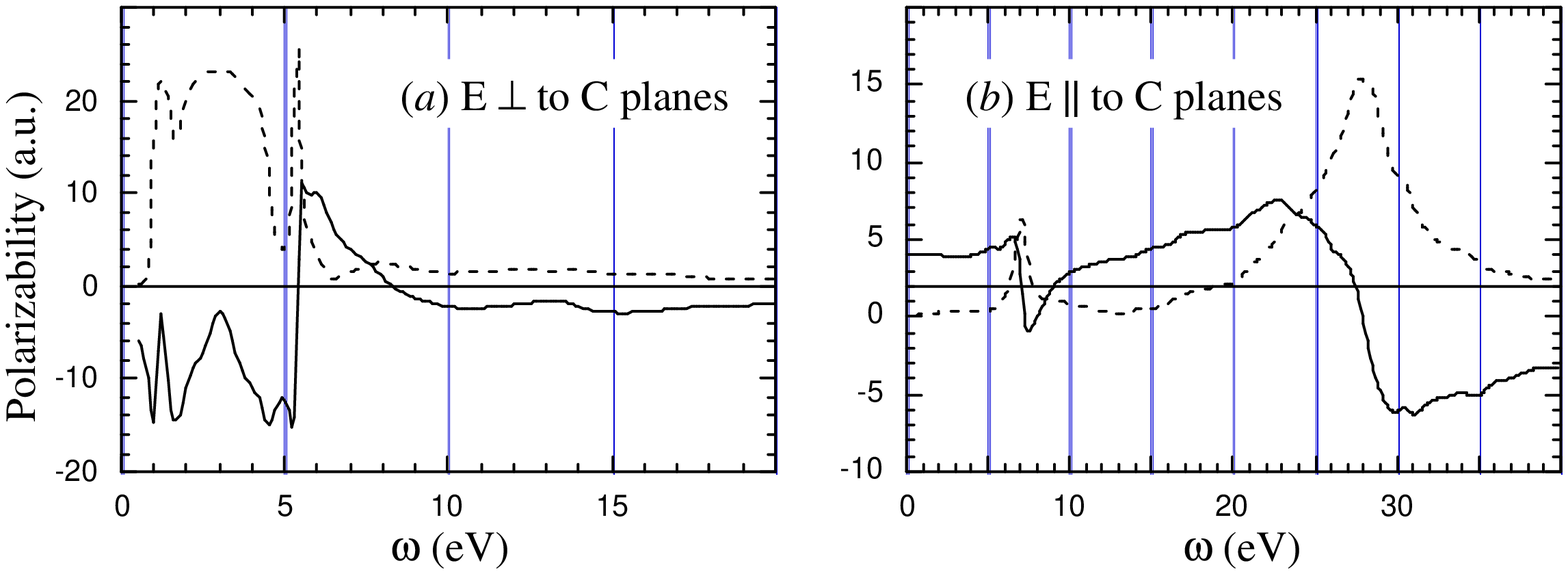}}}
\caption{Polarizability of a carbon atom in graphite for the
electric field perpendicular (a) and parallel (b) to the atomic
planes.} \label{Polarizability}
\end{figure}

In order to apply this theory to fullerenes, one must obtain first
the polarizability of the C atoms in the molecule. This is done
via the Clausius-Mosotti relation by using the bulk dielectric
function of graphite, \cite{palik} assuming that the
polarizability of C atoms is essentially the same in graphite and
in fullerenes owing to their similar local atomic environment. In
Fig.\ \ref{Polarizability}, we plot both the real and the
imaginary parts of the two independent components of the
polarizability tensor (i.e., normal and parallel to the graphite
planes). These plots correspond to  C atoms in a graphite-like
structure and have no physical meaning unless the coupling with
neighboring atoms is considered. It is interesting to note that
both components present resonances around $6$ eV, which lead to
the excitation usually referred to as $\pi$, although the coupling
with the $\sigma$ electrons in the plane is also important, as
shown by the actual position of this peak in the EELS data
discussed below. On the other hand, the broad peak of the
imaginary part of the parallel component around $27$ eV reproduces
correctly the loss spectrum of graphite. \cite{Saito91} This
correspondence between some polarizability features and EELS
experiments reflects the fact that $\sigma$ electrons are almost
uncoupled at energies above 10 eV in graphite: the normal
component of the polarization is very small and the coupling
between $\sigma$ electrons within a carbon plane vanishes due to
the symmetry of the lattice (the coupling between atoms in
different planes is also very weak due to the large separation
between planes). Nevertheless, the surface curvature in carbon
nanostructures breaks that symmetry, and therefore, the spectra
become sensitive to the geometrical shape of the sample over
distances larger than the interatomic separation.

\begin{figure}
\centerline{\scalebox{0.36}{\includegraphics{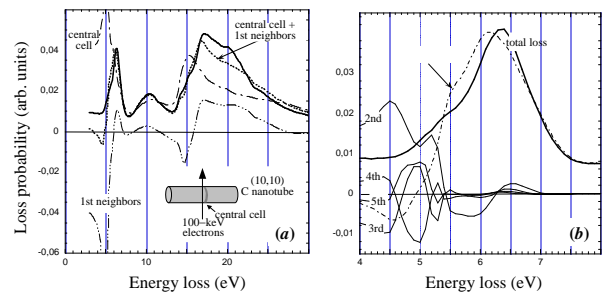}}}
\caption{{\bf (a)} Energy loss probability per carbon atom in a
$(10,10)$ carbon nanotube. The primary electron energy is 100 keV
and the collector aperture half-angle is $\theta_{c}=1$ mrad. The
solid curve represents the total energy loss probability, while
the rest of the curves stand for the partial contribution of the
central cell, the first neighboring cells, and the central cell
plus the first neighboring cells (see labels and discussion in the
text). {\bf (b)} Detail of the low-energy region of the spectrum
in (a), showing separate contributions of up to the fifth
neighboring cells.} \label{tube1}
\end{figure}

Let us first study the energy loss of an electron beam in a
$(10,10)$ carbon nanotube to illustrate a case of an infinite
atomic structure. \cite{Lin94,Dunlap94} The basic cell of this
tube consists of a ring of diameter $1.37$ nm, formed by $40$ C
atoms. This ring is periodically repeated with a period of
$d=0.247$ nm along the tube axis.  For each C atom, the
polarizability tensor has been built so that its normal axis is
perpendicular to the plane formed by the three closest atoms. We
have considered a typical experiment in transmission electron
microscope (TEM), where a $100$-keV electron beam is impinging
normal to the cylinder axis. The screened interaction has been
calculated in momentum space as shown in Sec.\ \ref{ap-a}, and we
work out the energy loss probability per C atom by taking into
account the contribution of neighboring cells to the sum of
expression (\ref{pqw1}) [ten cells at each side of the central
cell are enough to get good convergence above $3$ eV (see Fig.\
\ref{tube1}(b))]. Fig. \ref{tube1}(a) shows the energy loss
probability for a typical value of the collector aperture of
$\theta_{c}=1$ mrad. The spectrum exhibits three features around
$6.5$, $11$, and between $17-20$ eV, which according to the
analysis of Ref. [\onlinecite{Reed01}] can be identified as
$\pi$-surface plasmons and both low- and high-energy $\sigma+\pi$
surface plasmons, respectively. Below the $\pi$ feature, there is
a shoulder in the loss spectrum that has been experimentally
observed in both fullerenes
\cite{Gensterblum91,lucas92,Keller92,Kuzuo94,Kuzuo95} and
nanotubes. \cite{Reed01,Liu02} Comparison with Fig.\
\ref{Polarizability} permits to postulate that these peaks derive
from the coupling between the resonances of the $\pi$ and $\sigma$
electrons.

In Fig. \ref{tube1}(a), we have also plotted the contributions to
the sum of Eq. (\ref{pqw1}) of the first neighboring cells with
respect to a given central cell. Although the coupling between
different cells involves atoms at different distances from each
individual atom, these plots provide valuable information on the
distance dependence of the electrodynamic interaction between
atoms. First, one can see that the central cell and the first
neighboring cells are give contributions of the same order of
magnitude. The strong cancellation between contributions from the
central cell and the first neighbors at $\omega=5$ eV (i.e., the
position of the sharp resonance in the normal component of the
polarizability) proves the collective nature of $\pi$-plasmon
resonances. This plot shows that the coupling between nearest
cells gives rise almost entirely the total energy loss probability
between $7$ eV and $17$ eV, but more distant cells are needed
outside this energy range. This point is relevant to the study of
the $\sigma+\pi$ plasmon, which is around $24$ eV in fullerenes,
whereas it shows up around $16$ eV in nanotubes. As discussed
above, this energy shift from fullerenes to nanotubes can be
explained by the influence of curvature on the coupling between
$\sigma$ orbitals. In Fig.\ \ref{tube1}(b), we plot a detail of
the low-energy region of the same spectrum, where the contribution
of more neighboring cells has been plotted separately. The
$\pi$-plasmon peak is basically determined by coupling of the
central cell with first neighbors, while the contribution of more
distant cells overcomes that of first neighbors below this peak.
This observation suggests that the position of the $\pi$-plasmon
feature depends very little on the long range structure of the
sample, while the sub $\pi$-plasmon features are more sensitive to
geometry on a larger scale.

\begin{figure}
\centerline{\scalebox{0.36}{\includegraphics{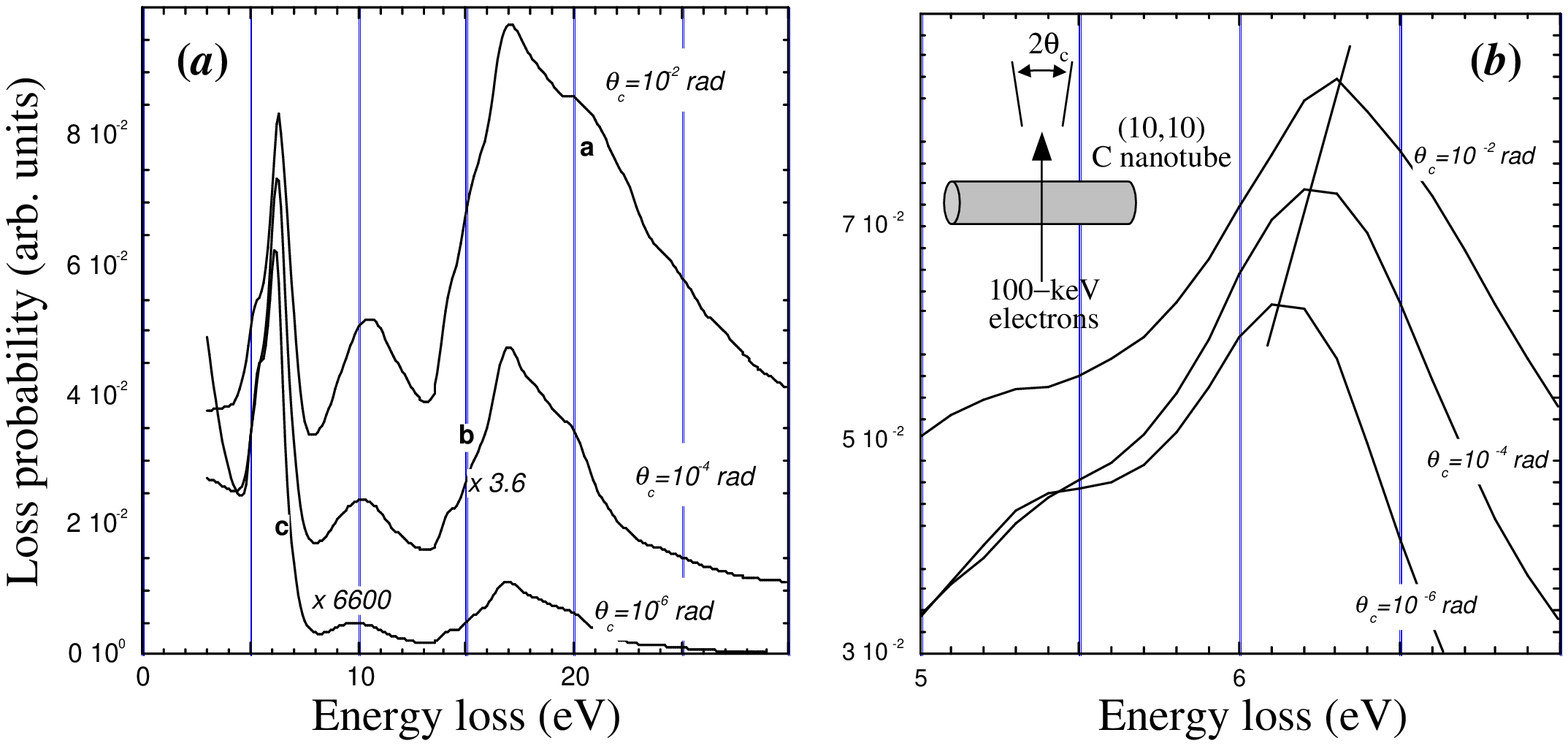}}}
\caption{{\bf (a)} Energy loss probability per carbon atom in a
$(10,10)$ carbon nanotube for three different collector aperture
half-angles (see labels) . The primary electron energy is 100 keV.
{\bf (b)} Detail of the $\pi$-plasmon peak in (a) showing a
positive dispersion with increasing aperture size.} \label{tube2}
\end{figure}

Reed and Sarikaya \cite{Reed01} have measured loss spectra
corresponding to a single nanotube of about $1.2$ nm in diameter
using a $100$-keV beam focused over a region of $0.2$ nm in
diameter. Their sample is similar to the one used in our
theoretical simulations. The spectrum corresponding to the probe
passing through the tube presents a broad resonance around $17$
eV, while a faint signature of a low energy resonance is hidden by
the elastic peak. For large impact parameters (several nanometers
away from the tube axis), the low-energy loss peak becomes sharper
and the position of the $\sigma+\pi$ plasmon feature is red
shifted to $15$ eV. Although the comparison of these data with
those derived from our model is not straightforward because of the
different type of beam, one can claim good qualitative agreement
between them. The contribution of electrons with large impact
parameter can be selected by collecting the electrons scattered
within a small scattering angle. In Fig.\ \ref{tube2}(a) we show
three spectra corresponding to different values of the collector
aperture $\theta_{c}$. As the angle $\theta_{c}$ decreases, the
$\pi$ peak becomes relatively higher, while the center of the
$\sigma+\pi$ peak shifts down in energy. This dependence of the
intensity of the different peaks of the spectrum on the scattering
angle has been also experimentally confirmed for C$_{60}$.
\cite{Keller92} A simple explanation based on the classical EELS
theory has been proposed by Reed and Sarikaya,\cite{Reed01}
namely, that the impact parameter dependence of the energy loss
probability goes like
\begin{equation}
P(\omega)\sim e^{-2\frac{\omega d}{v}},\hspace{1cm} d \geq
v/\omega, \label{spher}
\end{equation}
where $d$ is the distance relative to a given scattering center.
\cite{Ferrell85} This means that distant collisions (i.e., small
momentum transfers) excite low-energy modes more efficiently, a
fact that explains the positive dispersion of the loss peaks. The
analysis of the $\pi$-plasmon feature of Fig.\ \ref{tube2}(b) also
shows a weak positive dispersion, which is in qualitative
agreement with momentum resolved EELS measurements in multishell
fullerenes and peapods. \cite{Pichler01,Liu02}

\begin{figure}
\centerline{\scalebox{0.36}{\includegraphics{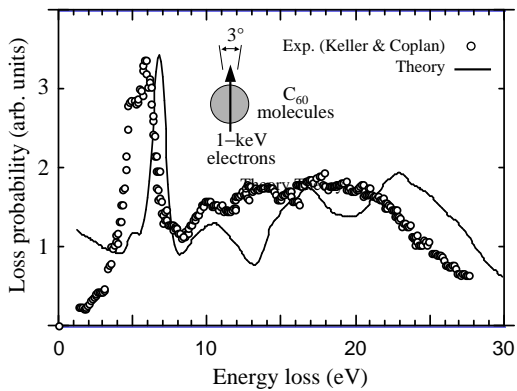}}}
\caption{Comparison between theory (solid curve) and experiment
(circles, taken from Ref. [5]) for the energy loss probability of
$1$-keV electrons passing through a C$_{60}$ sample in the gas.
The collector aperture half-angle is $1.5^\circ$.} \label{Keller}
\end{figure}

\begin{figure}
\centerline{\scalebox{0.36}{\includegraphics{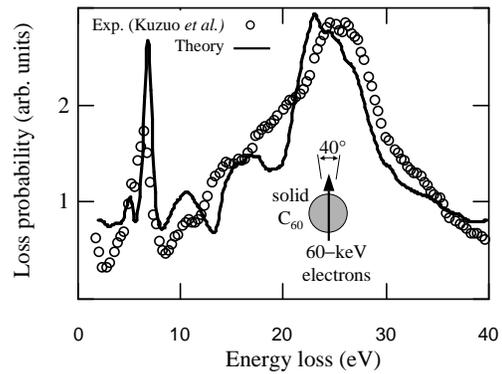}}}
\caption{Comparison between theory (solid curve) and experiment
(circles, taken from Ref. [11]) for the energy loss probability of
$60$-keV electrons in solid C$_{60}$. The collector aperture
half-angle (not specified in the experimental reference) has been
set to $20^\circ$.} \label{Kuzuo}
\end{figure}

\begin{figure}
\centerline{\scalebox{0.36}{\includegraphics{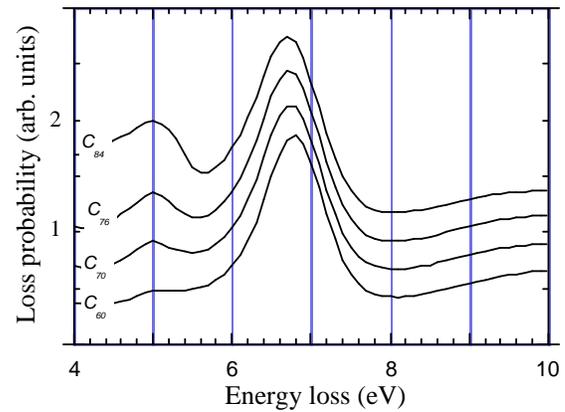}}}
\caption{Electron energy loss spectra calculated for different
spherical carbon molecules. The plots have been shifted upwards to
allow comparison. The primary electron energy is 60 keV. Electrons
are collected up to a maximum scattering angle of $1$ mrad.}
\label{c60series}
\end{figure}

The EELS simulation corresponding to spherical fullerenes is more
straightforward, since the system has now a finite number of C
atoms. \cite{datac60} In Fig.\ \ref{Keller}, we show the resulting
computed loss spectrum in C$_{60}$ as compared to experimental
data reported by Keller and Coplan, \cite{Keller92} measured in
the gas phase, so that, as discussed in Sec.\ \ref{Introduction},
our simulation based upon a broad beam configuration suits best
the experiment. The energy of the electrons is $1$ keV ($v=8.6$
a.u) and the collector aperture is $1.5^{\circ}$, which is small
enough to justify the neglect of recoil, even at such a relatively
low velocities. Although the theoretical spectrum seems to be
shifted up in energy, the simulation reproduces reasonably well
the relative position of the peaks (including the small sub $\pi$
plasmon structure) as well as their relative intensity. In Fig.
\ref{Kuzuo}, we compare the results of our theory to the
experiments of Kuzuo and coworkers \cite{Kuzuo95} for crystalline
C$_{60}$. The probe energy is now $60$ keV ($v=61$ a.u.). We
obtain good agreement in this case as well, suggesting that the
coupling between neighboring target fullerenes is weak. This
spectrum is very similar to those corresponding to C$_{70}$,
C$_{76}$, and C$_{84}$, with only small differences in the low
energy region below the $\pi$-plasmon feature, as shown in Fig.\
\ref{c60series}. The $\pi$ plasmon around $6.7$ eV is very similar
in all cases, except for a small red shift as the number of atoms
in the molecule increases, in agreement with experiments using
crystalline fullerenes, \cite{Kuzuo95} where the $\pi$ peak lies
in the $6.4-6.1$ eV interval. (Nevertheless, one has to keep in
mind that the interaction between neighboring fullerenes in the
solid phase can play a role that is not described by the present
calculations.) The almost identical shape of the $\pi$-plasmon
peak in this series (Fig.\ \ref{c60series}) is in contrast with
the strong differences at lower energy losses, as a consequence of
the fact that the local atomic structure of all of these molecules
is very similar up to first neighbors (which give the dominant
contribution to the $\pi$ plasmon), while differences in
geometrical ordering over larger distances play a significant role
at lower energies.

In summary, we have obtained the screened interaction of a system
of polarizable atoms to an external electric field and have
applied this function to study the energy loss experienced by an
electron (within a fully quantum-mechanical treatment of the
probe) passing near carbon fullerenes and nanotubes. The spatial
range of the coupling between dipoles explains the sensitivity of
EELS to the structural details of the carbon nanostructures. The
good agreement between theory and experiments is encouraging
regarding the applicability of the derived screened interaction to
other problems involving geometrically well-defined atomic
structures.


\acknowledgments

This work has been supported in part by the Basque Departamento de
Educaci\'{o}n, Universidades e Investigaci\'{o}n, the University
of the Basque Country UPV/EHU (contract No. 00206.215-13639/2001),
and the Spanish Ministerio de Ciencia y Tecnolog\'{\i}a (contract
No. MAT2001-0946).



\end{document}